\documentclass{llncs}
\usepackage[utf8]{inputenc}
\usepackage[T1]{fontenc}
\usepackage{amsmath,amssymb}
\usepackage{graphicx}
\usepackage{xcolor}

\newcommand{\squishlist}{
 \begin{list}{$\bullet$}
  { \setlength{\itemsep}{0pt}
     \setlength{\parsep}{1pt}
     \setlength{\topsep}{1pt}
     \setlength{\partopsep}{0pt}
     \setlength{\leftmargin}{1.5em}
     \setlength{\labelwidth}{1em}
     \setlength{\labelsep}{0.5em} } }
 \newcommand{\squishend}{\end{list}}

\title{Distributed Processing of Generalized Graph-Pattern Queries in SPARQL 1.1}
\author{Sairam Gurajada\inst{1} \and Martin Theobald\inst{2}}
\institute{Max-Planck-Institute for Informatics, \email{gurajada@mpi-inf.mpg.de} \and University of Ulm, \email{martin.theobald@uni-ulm.de}}

\begin{document}
\maketitle

\begin{abstract}
We propose an efficient and scalable architecture for processing generalized graph-pattern queries as they are specified by the current W3C recommendation of the SPARQL 1.1 ``Query Language'' component. Specifically, the class of queries we consider consists of sets of SPARQL triple patterns with labeled property paths. From a relational perspective, this class resolves to conjunctive queries of {\em relational joins} with additional {\em graph-reachability predicates}. 
For the scalable, i.e., distributed, processing of this kind of queries over very large RDF collections, we develop a suitable partitioning and indexing scheme, which allows us to shard the RDF triples over an entire cluster of compute nodes and to process an incoming SPARQL query over all of the relevant graph partitions (and thus compute nodes) in parallel. Unlike most prior works in this field, we specifically aim at the {\em unified optimization and distributed processing} of queries consisting of both relational joins and graph-reachability predicates. All communication among the compute nodes is established via a proprietary, asynchronous communication protocol based on the Message Passing Interface.
\end{abstract}

\vspace{-2mm}
\section{Introduction}
\label{sec:intro}
\vspace{-2mm}
\subsection{Background \& Motivation}
\vspace{-2mm}
RDF~\cite{rdfs} and SPARQL~\cite{sparql} are two standards recently recommen\-ded by the W3C for representing and querying linked data on the Web. RDF is a very versatile data format and hence found a wide adoption in many communities. 
Due to its broad applicabi\-lity, it is frequently employed as a generic, albeit simple, know\-ledge representation format for large-scale information-extraction endeavors such as DBpedia, Freebase or YAGO, for capturing relationships in social networks, and many others.
Consequently, much recent work has focused on the distributed, and hence scalable, processing of {\em graph-pattern queries} in SPARQL 1.0 via a variety of SQL~\cite{Sakr:2010:RPR:1815948.1815953,monetdb} and NoSQL architectures~\cite{DBLP:conf/semweb/Cudre-MaurouxEFGHHKMSW13,DBLP:conf/btw/HagedornHS15}. Engines such as SHARD~\cite{shard}, SW-Store~\cite{mrrdf3x} (which is a Hadoop extension of the centralized RDF-3X~\cite{rdf3x10} engine), EAGRE~\cite{eagre}, Trinity.RDF~\cite{trinity.rdf}, our own TriAD~\cite{triad} engine, as well as commercial tools like Virtuoso~\cite{virtuoso} and Ontobroker~\cite{ontobroker} evaluate SPARQL queries either as series of relational joins~\cite{triad,mrrdf3x,shard,eagre} or via iterative graph exploration and message passing~\cite{trinity.rdf}. 

With its recent update, 1.1, SPARQL underwent a major revision. From a relational perspective, the introduction of {\em property paths}~\cite{propertypaths} likely constitutes the most remarkable change. Property paths allow for annotating pairs of query vertices by regular expressions in which properties and entire paths may be marked by a Kleene ``{\tt +}'' or ``{\tt *}''. SPARQL 1.1 thus introduces a notion of {\em generalized graph-pattern queries} in which property paths express transitive reachability constraints among sets of RDF entities that become bound to the query variables.
That is, unlike in SPARQL 1.0, joins among triple patterns may be recursive and need to be evaluated either iteratively, or they involve a potentially costly materialization of the properties' closures.

As opposed to the large variety of SPARQL 1.0 engines, the processing of property paths in SPARQL 1.1
so far has been investigated by only very few approaches~\cite{virtuoso,DBLP:conf/sigmod/GubichevBS13,DBLP:conf/esws/Przyjaciel-ZablockiSHL11} (of which only~\cite{virtuoso} is available). A particular challenge lies in the combined optimization of relational joins among the triple patterns (based on shared variables) with additional graph-reachability predicates (based on one or more properties marked by a ``{\tt +}'' or ``{\tt *}''). While state-of-art indexing techniques for graph-reachability predicates~\cite{DBLP:conf/www/GaoA13,DBLP:conf/icde/SeufertABW13,grail} are inherently limited to a centralized setting, we are currently aware of only one approach that specifically tackles distributed reachability queries for single-source, single-target queries ~\cite{DBLP:journals/pvldb/FanWW12}. For multi-source, multi-target reachability queries, as they frequently occur in SPARQL 1.1, we are aware of just two centralized approaches~\cite{DBLP:conf/www/GaoA13,DBLP:journals/pvldb/ThenKCHPK0V14} that provide suitable indexing and processing strategies. 
Distributed graph engines, such as Berkeley's GraphX~\cite{graphx}  and Apache Giraph, on the other hand, allow for the scalable processing of graph queries over massive, partitioned data graphs. Both provide generic API's for implementing various kinds of graph algorithms, including multi-source, multi-target reachability queries. However, they do not support the kinds of indexing techniques known from the centralized approaches, and they are not directly amenable to the declarative way of querying RDF data as it is suggested by the SPARQL standard. Processing property paths with transitive reachability constraints under these frameworks requires an iterative form of graph traversal, which may result in as many iterations (and hence communication rounds) as the diameter of the graph in the worst case.

Summarizing this motivating section, we are not aware of any previous approach that achieves a true scale-out in processing generalized graph-pattern queries for SPARQL 1.1 under both {\em strong} (thus reducing the query time by increasing the number of processors) and {\em weak scaling} (thus achieving predictable query times when increasing both the data size and the number of processors). In this paper, we present a distributed architecture that takes advantage of indexing techniques for graph-reachability predicates known from centralized approaches, while retaining the declarative style of querying massive RDF data graphs for SPARQL 1.1 with property paths.
\vspace{-2mm}
\subsection{Contributions}
\vspace{-2mm}
\mbox{We summarize the contributions of our work as follows.}
\squishlist
\item We present an {\em efficient} and {\em scalable query engine} for the core of the SPARQL 1.1. Specifically, the fragment of SPARQL we consider resolves to sets of triple patterns with labeled property paths, which allows for formulating generalized graph pattern queries as conjunctive queries of relational joins with additional graph-reachability predicates (including one or more properties marked by a Kleene ``{\tt +}'' or ``{\tt *}'').
\item We provide a {\em unified indexing scheme},  {\em cost model} and {\em query optimization framework} to seamlessly integrate graph-reachability predicates into the relational query processor of our TriAD~\cite{triad} engine. TriAD$^*$ employs a strictly fixed, asynchronous message-passing protocol to evaluate a SPARQL 1.1 query among all of the compute nodes in parallel. Our protocol requires exactly one round of communication per graph-reachability predicate and thus avoids a costly, iterative form of communication.
\item Our approach is the first to report an actual {\em horizontal scale-out} in processing SPARQL 1.1 queries with property paths over a number of large RDF collections. We present a detailed experimental evaluation of our approach under both strong and weak scaling, and in comparison to the Virtuoso native RDF store.
\squishend


\vspace{-2mm}
\section{RDF Data \& SPARQL Query Model}
\label{sec:prelim}
\vspace{-2mm}
We next formally define our data and query model. Our model is based on a graph representation of RDF data and supports the core syntax of SPARQL 1.1 to express queries over an RDF data graph as graph-pattern queries. 
Specifically, we focus on the {\tt WHERE} clause of a SPARQL 1.1 query, in which triple patterns (including triples annotated with property paths) are connected (i.e., ``joined'') via their shared variables. We currently omit other SPARQL 1.0/1.1 extensions such as {\tt UNION}, {\tt FILTER}, aggregations, subqueries and negation. 

\vspace{-2mm}
\subsection{RDF Data Model}
\label{sec:datamodel}
\vspace{-2mm}
We follow the general W3C recommendation of the basic RDF vocabulary~\cite{rdfs} (however ignoring RDF schema extensions), in which an RDF collection consists of a {\em set of triples} which are each of the form $\langle\mathit{subject}$, $\mathit{property}$, $\mathit{object}\rangle$ (or $\langle s$, $p$, $o\rangle$, for short). 
A set of RDF triples can thus concisely be represented as a directed, labeled multi-graph, as it is defined next.
\begin{definition}
\label{def:rdf}
An \textbf{RDF data graph} $G_D(V_D,E_D,\mathit{Cons},$ $\phi_D)$ is a directed, labeled multi-graph where $V_D$ is the set of data vertices, $E_D$ is the set of directed edges connecting the vertices in $V_D$, $\mathit{Cons}$ is the set of vertex and edge labels, and $\phi_D$ is a labeling function with $\phi_D: V_D \cup E_D \rightarrow \mathit{Cons}$ s.t. $\forall v_i,v_j \in V_D$, $v_i \neq v_j$, it holds that $\phi_D(v_i) \neq \phi_D(v_j)$.
\end{definition}

\begin{figure}[htb]
\centering
\vspace{-6mm}
\scalebox{.3}{\includegraphics{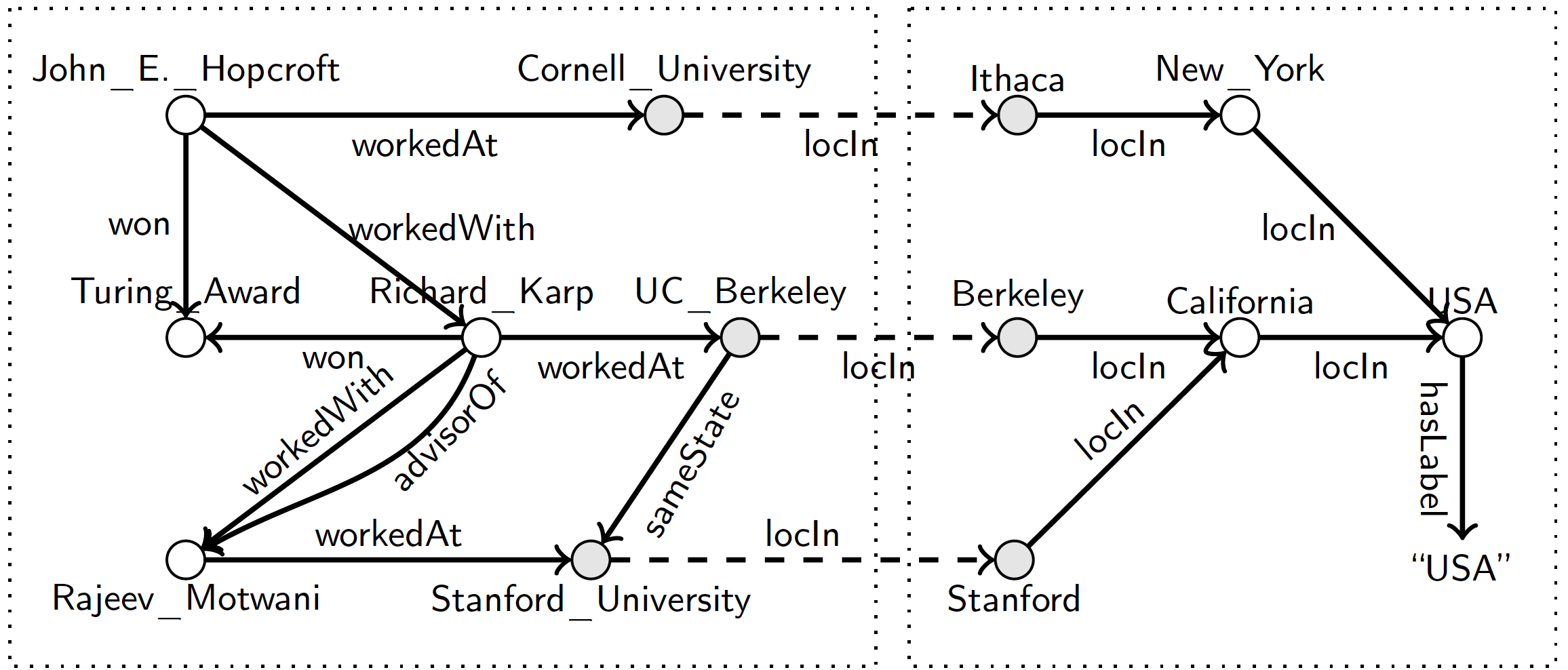}}
\vspace{-6mm}
\caption{A partitioned RDF data graph}
\vspace{-6mm}
\label{fig:rdfexample}
\end{figure}

\noindent\textbf{Partitioning RDF Data Graphs.} A potentially scalable approach for managing very large RDF collections is to follow a distributed indexing scheme. In our approach, an RDF data graph is partitioned into multiple graph partitions, such that one or more of these graph partitions can be managed locally by a compute node. More formally, an RDF data graph $G_D$ is partitioned into $k$ vertex- and edge-disjoint graphs $\mathcal{G}=$ $\{G_1, G_2,$ $ ..., G_k\}$, where each $G_i(V_i,E_i,\mathit{Cons},\phi_D)$ is a {\em vertex-induced subgraph} of $G_D$. That is, it holds that $\bigcup_{i=1..k} V_i = V_D$ and $V_i \cap V_j = \emptyset$ for all $i \neq j$. Figure~\ref{fig:rdfexample} shows an example RDF data graph that is partitioned into the two graph partitions $G_1$, $G_2$ which are located at slaves 1 and 2, respectively.

\vspace{-4mm}
\subsection{SPARQL Query Model}
\vspace{-2mm}
SPARQL 1.0~\cite{sparql} is a widely adopted W3C language recommendation for querying RDF data. A typical SPARQL 1.0 query comprises a {\em conjunction of triple patterns} of the form $\langle s$, $p$, $o\rangle$, where each of the $s$, $p$, $o$ components may denote either a constant or a variable. For instance, the query ``{\em Find all professors who won a Turing Award and worked in a US university}'' can be specified in SPARQL 1.0 as follows.

\vspace{2mm}
{
$\mathtt{SELECT~?person~WHERE~\{} $\\
$~~~~~~\mathtt{?person~won~Turing\_Award.~~?person~workedAt~?univ.}$\\ 
$~~~~~~\mathtt{?univ~locIn~?city.~~?city~locIn~?state.} $\\
$~~~~~~\mathtt{?state~locIn~?country.~~?country~hasLabel~"USA" \}}$
}
\vspace{2mm}

\noindent\textbf{Property Paths.} With the latest revision, SPARQL 1.1, {\em property paths}~\cite{propertypaths} were introduced to SPARQL 1.0. A property path specifies how an entity is (possibly transitively) connected to another entity. 
The above query can thus concisely be rewritten by using a property path as follows.

\vspace{2mm}
{
$\mathtt{SELECT~?person~WHERE~\{}$\\
$~~~~~~\mathtt{?person~won~Turing\_Award.}$\\
$~~~~~~\mathtt{?person~workedAt/locIn*/hasLabel~"USA"~\}}$
}
\vspace{2mm}

\noindent Here, the property path {${\mathtt{workedAt/locIn*/hasLabel}}$} expresses a transitive connectivity via the {\tt locIn} property. That is, a path connecting any vertex that becomes bound to the variable {\tt ?person} with the vertex with label ``$\mathtt{USA}$'' must involve zero or more consecutive properties with the label {\tt locIn}. 
Property paths thus help the user in two ways: (1) in simplifying the query representation and (2) in formulating queries with partial knowledge of the underlying schema.

\smallskip
\noindent\textbf{SwPP Queries.} We focus on a subset of queries expressible in SPARQL 1.1, in the following called ``SwPP'', consisting of {\em conjunctions of triple patterns with property paths}. 
Specifically, a triple pattern of an SwPP query again is of the form $\langle s$, $p$, $o\rangle$, where each of the $s$, $o$ components may refer to either a constant in $\mathit{Cons}$ or to a query variable from a distinct set of variables $\mathit{Vars}$. 
Moreover, $p$ may refer to a {\tt path} expression whose grammar we define below.

\begin{align}
\mathtt{path} &~\mathtt{:=}~ \mathtt{path} \mathtt{/} \mathtt{path} &(concatenation~of~paths) \\
     &~\mathtt{:=}~ \mathtt{URI} &(single~property)\\
     &~\mathtt{:=}~ \mathtt{\text{\^{}}URI} &(inverse~property~direction) \\
     &~\mathtt{:=}~ \mathtt{URI?} &(zero~or~one~property) \\
     &~\mathtt{:=}~ \mathtt{URI*} &(zero~or~more~properties) \\
     &~\mathtt{:=}~ \mathtt{URI+} &(one~or~more~properties)
\end{align}

Formally, an SwPP query thus again forms a directed, labeled multi-graph, as it is defined next.
\begin{definition}
\label{def:sparql}
A \textbf{SPARQL query graph} $G_Q$ $(V_Q,E_Q,\mathit{Cons},$ $\mathit{Vars},\phi_Q,\psi_Q)$ is a directed, labeled multi-graph where $V_Q$ is the set of query vertices, $E_Q$ is the set of edges connecting the vertices in $V_Q$, $\mathit{Cons}$ is the set of vertex and edge labels, $\mathit{Vars}$ is a set of query variables, $\phi_Q$ is a vertex-labeling function with $\phi_Q: V_Q \rightarrow \mathit{Cons} \cup \mathit{Vars}$, and $\psi_Q$ is an edge-labeling function with $\psi_Q: E_Q \rightarrow \{\mathtt{\text{\^{}}}\}\,\mathtt{|}\,\mathit{Cons}\,\mathtt{|}\,\{\mathtt{*},\mathtt{+}, \mathtt{?}\}$.
\end{definition}

We hereby adopt a simpler definition for property paths than the full syntax proposed by the W3C~\cite{propertypaths}. However, by rewriting an entire path expression (denoted as ``{$\mathtt {path}$}'' in the above grammar) into a sequence of join conditions, each with a property that denotes a single URI (referred to as ``{$\mathtt {URI}$'') with an optional Kleene ``{\tt *}'' or ``{\tt +}'', we allow a more general syntax for property paths rather than just a single transitive property. The above grammar in particular allows concatenations of properties into paths of arbitrary length, as long as these can be rewritten into a conjunction of triple patterns and thus conform to Definition~\ref{def:sparql}. In our implementation, the distinction between ``{\tt +}'', ``{\tt *}'', ``{\tt ?}'' and ``{\tt \^{}}'' is very simple. For ``{\tt +}'', we merely disallow an equality between a source and a target vertex; while for ``{\tt ?}'', we restrict the maximum path length to 1. The inversion ``{\tt \^{}}'' simply swaps the source and target vertices.

\vspace{-2mm}
\section{Architecture}
\label{sec:architecture}
\vspace{-2mm}
The architecture of our SPARQL 1.1 engine (in the following coined TriAD*) is based on TriAD \cite{triad,DBLP:conf/sigmod/GurajadaSMT14a}, which we originally developed for processing conjunctions of triple patterns in SPARQL 1.0. The design of TriAD in principle follows a classical master-slave architecture at indexing time, but allows for a direct, asynchronous communication among the slaves at query-processing time.

\smallskip
\noindent\textbf{Indexing.}  At indexing time, the master node is responsible for the encoding of incoming RDF triples and for the distribution of these triples among the slave nodes. The slaves then construct their local index structures (each over a distinct graph partition) and send their local index statistics back to the master node. 

\smallskip
\noindent\textbf{Query Processing.} At query time, the master node facilitates these precomputed index statistics to compile a global query plan that is then distributed to and processed at all slaves in parallel. While processing such a query plan, the slaves directly exchange their intermediate query results via a proprietary, asynchronous communication protocol based on the Message Passing Interface~\cite{mpi}.

\smallskip
\noindent To facilitate the processing of SPARQL 1.1 queries with property paths, TriAD* augments the architecture of TriAD in two decisive ways:
\squishlist
\item {\bf Triple \& Reachability Indexes:} 
We create two kinds of index structures---one that is optimized for processing joins and another one that is optimized for graph-reachability queries among the slaves (see Section~\ref{sec:indexing}). 
\item {\bf Unified Query Optimization:} We devise a unified cost model for optimizing both relational joins and graph-reachability predicates (see Section~\ref{sec:optpro}). 
\squishend

\vspace{-2mm}
\section{Index Organization}
\label{sec:indexing}
\vspace{-2mm}



We use a hashing-based sharding scheme to distribute RDF triples across the slaves~\cite{triad,mrrdf3x,shard}. Each RDF triple in a collection is parsed at the master node and encoded into an integer format via a dictionary to obtain compact identifiers for each of the $s$, $p$ and $o$ components. Every such encoded $\langle s$, $p$, $o\rangle$ triple is then distributed to (at most) two slaves $i$, $j$ by choosing $i =(s \mod k)$ and $j = (o \mod k)$ as sharding conditions, respectively. Although we use a hash function for sharding, we can easily adopt different partitioning schemes, for example, by exchanging $s$, $o$ in the hashes with the graph partitions to which $s$, $o$ are assigned by a locality-preserving graph-partitioning framework such as METIS~\cite{Karypis:1998:FHQ:305219.305248}.

\vspace{-4mm}
\subsection{Triple Indexes}
\label{sec:tripleindexes}
\vspace{-2mm}
Following TriAD's general indexing strategy~\cite{triad}, each slave creates six permutations of in-memory triple indexes for the efficient processing of relational joins among triple patterns based on their subjects or objects. 
Depending on whether an incoming triple was hashed to slave $i$ via its subject or object, these six permutations are arranged into two groups: (i) the three {\it subject-key} indexes (SPO, SOP, PSO) and (ii) the three {\it object-key} indexes (OSP, OPS, POS). That is, a sharded triple, which was hashed onto slave $i$ via its subject field, is indexed in three permutations by using the subject as key. Likewise, a triple hashed onto slave $i$ via its object field is indexed three times by using the object as key. The two PSO and POS permutations are employed for index scans of triple patterns with a given property and an optional subject or object as key.

\smallskip
\noindent\textbf{Sorting \& Scans.} The six vectors storing the SPO permutations at each slave are each sorted with respect to their permutation order. For example, the OPS vector is sorted primarily by the object field, followed by the predicate field, and finally by the subject field. Thus, when using a given object as search key, we first determine the entry point in the OPS vector via a binary search and then start scanning all triples under this search key sequentially.

\vspace{-4mm}
\subsection{Indexes for Property Paths}
\label{sec:swppindexes}
\vspace{-2mm}
In addition to the triple indexes, each slave constructs a separate reachability index to facilitate the efficient processing of property paths in SPARQL 1.1. One such reachability index is built for each unique property label $p$ in the RDF collection, 
similar to the approach described in~\cite{DBLP:conf/sigmod/GubichevBS13}. The construction of our reachability index for a property $p$ over a partitioned RDF graph is based on our recent results from~\cite{DBLP:conf/sigmod/GurajadaT16}, which we briefly summarize below.

Let $\mathcal G^p=\{ G_1^p,G_2^p,\ldots,G_k^p\}$ be the $p$-induced subgraphs of an RDF data graph $G_D$. That is, each $G^p_i$ is a subgraph of a previously determined graph partition $G_i$ (e.g., by using METIS~\cite{Karypis:1998:FHQ:305219.305248} to partition the original multi-labeled data graph $G_D$) that comprises of only $p$-labeled edges. For each such subgraph $G_i^p$, we first identify the set of vertices $I_i^p$ (coined ``{\em in-boundaries}'') and $O_i^p$ (coined ``{\em out-boundaries}'') that lie on the cut $C$ that is given by $\mathcal G^p$. 
Next, for each partition $G_i^p$, we precompute all reachable pairs of vertices from in-boundaries $I_i^p$ to out-boundaries $O_i^p$. We then construct two graph-based index structures, called {\em boundary graph} and {\em compound graph} (described in detail in~\cite{DBLP:conf/sigmod/GurajadaT16}), as follows. 
Each set of reachable pairs for a partition $G_i^p$ is represented as a bipartite graph with edges from in-boundaries to out-boundaries that are transitively reachable. Boundary nodes which are both in- and out-boundaries are omitted from this step. The bipartite graphs for each partition $G_i^p$ are then communicated to (i.e., replicated across) all slaves $j$ (for all $j \neq i$). 
At each slave $i$, a {\it boundary graph} $B_i^p$ is constructed by merging the received bipartite graphs from slaves $j$ (for all $j \neq i$) with the cut $C$. Finally, we construct a compound graph $C^p_i$ by merging the boundary graph $B^p_i$ also with the local subgraph $G^p_i$.

\smallskip
\noindent\textbf{Reachability Indexes.} To reduce the sizes of the precomputed boundary graphs $B_i^p$, on the one hand, and to further accelerate graph-reachability queries over the compound graphs $C_i^p$, on the other hand, we apply the following optimizations (also described in detail in~\cite{DBLP:conf/sigmod/GurajadaT16}). First, we determine equivalence sets over in- and out-boundaries $I^p_i$ and $O^p_i$. The in- and out-boundaries are then redefined with respect to new virtual vertices, each representing an equivalence set. That is, $I^p_i$ and $O^p_i$ comprises of all {\it in-virtual} and {\it out-virtual} vertices, respectively. The bipartite graphs are then computed over the redefined $I^p_i$ and $O^p_i$ sets. This technique helps to reduce the size of the boundary graphs $B^p_i$. The compressed boundary graphs are then used to construct the compound graphs $C^p_i$ at each slave $i$ in the same manner as before. Second, the obtained compound graphs are each condensed into a {\em directed acyclic graph} (DAG), in which each vertex represents a strongly connected component (SCC) of the compound graph. We finally store these DAGs as in-memory adjacency lists (serialized into another pair of PSO and POS vectors) and perform a local DFS traversal to answer a graph-reachability query at each slave.

\vspace{-4mm}
\subsection{Index Statistics}
\vspace{-2mm}
To optimize SwPP queries consisting of both relational joins among triple patterns and of triple patterns with property paths, we rely on a cost-based plan generator which is part of TriAD's architecture. For this, we collect various statistics over the RDF data for both the basic triple patterns and triple patterns with property paths.

\smallskip
\noindent\textbf{Statistics for Triple Patterns.} As in~\cite{triad}, our statistics for basic triple patterns include (i) cardinalities $\mathit{Card(R_i)}$ of relations $R_i$ induced by individual {\em subject}, {\em property} and {\em object} keys and (ii) of relations induced by {\em subject}-{\em object}, {\em property}-{\em subject} and {\em property}-{\em object} pairs. In addition, we compute the join selectivities $\mathit{Sel(p_i,p_j)}$ of all {\em pairs of properties} $p_i$, $p_j$ to estimate the cardinality of a join among two triple patterns. 

\smallskip
\noindent\textbf{Statistics for Property Paths.} In order to plug triple patterns with property paths into our optimizer, we need to also estimate the selectivity of a property path. Precomputing these selectivities for every possible property path that may occur in a query is clearly intractable. We thus follow a simple sampling-based approach. For each individual property $p$, 
we determine the {\em reachability selectivity} $\mathit{Sel}(p)$ as the fraction of randomly sampled source and target vertices $(s,t)$, for which $s\leadsto t$ holds with respect to the subgraph of $G_D$ that is induced by $p$.



\vspace{-4mm}
\section{Query Optimization \& Distributed Processing}
\label{sec:optpro}
\vspace{-2mm}
We first translate SwPP queries in SPARQL 1.1 into a graphical representation, which then forms the basis for query optimization. These query graphs are generated by introducing a vertex for each triple pattern in the query, while the edges that connect two such vertices represent equi-joins. These equi-joins are based on the variables (i.e., either the subjects or objects) that are shared by two such triple patterns. Edges for equi-joins are labeled with the shared variables.
In addition to SPARQL 1.0 queries, SwPP queries contain triple patterns with property paths. Following~\cite{DBLP:conf/sigmod/GubichevBS13}, we represent a property path by a distinguished edge among two such query vertices, whose labels denote the reachability predicates among the subjects or objects in the respective vertices' triple patterns. In case the subject or object of a connected triple pattern is either a constant or an unbound variable (i.e., the variable is not present in the other triple patterns), we create a new query vertex for the same and add an edge between the respective query vertices. 

\begin{figure}[htb]
\vspace{-4mm}
\begin{minipage}[t]{0.5\textwidth}
\hspace*{4mm}{\tt SELECT * WHERE \{}\\
\hspace*{4mm}{\tt P$_1$:\hspace*{3mm} ?p workedAt ?u.}\\
\hspace*{4mm}{\tt P$_2$:\hspace*{3mm} ?p won "Turing\_Award".}\\
\hspace*{4mm}{\tt P$_3$:\hspace*{3mm} ?p1 workedAt ?u1.}\\
\hspace*{4mm}{\tt P$_4$:\hspace*{3mm} ?u locIn* "USA".}\\
\hspace*{4mm}{\tt P$_5$:\hspace*{3mm} ?p workedWith* ?p1.}\\
\hspace*{4mm}{\tt P$_6$:\hspace*{3mm} ?u sameState* ?u1.\}}\\
\end{minipage}
\hfill
\begin{minipage}[t]{0.4\textwidth}
\vspace{-4mm}
\scalebox{.145}{\includegraphics{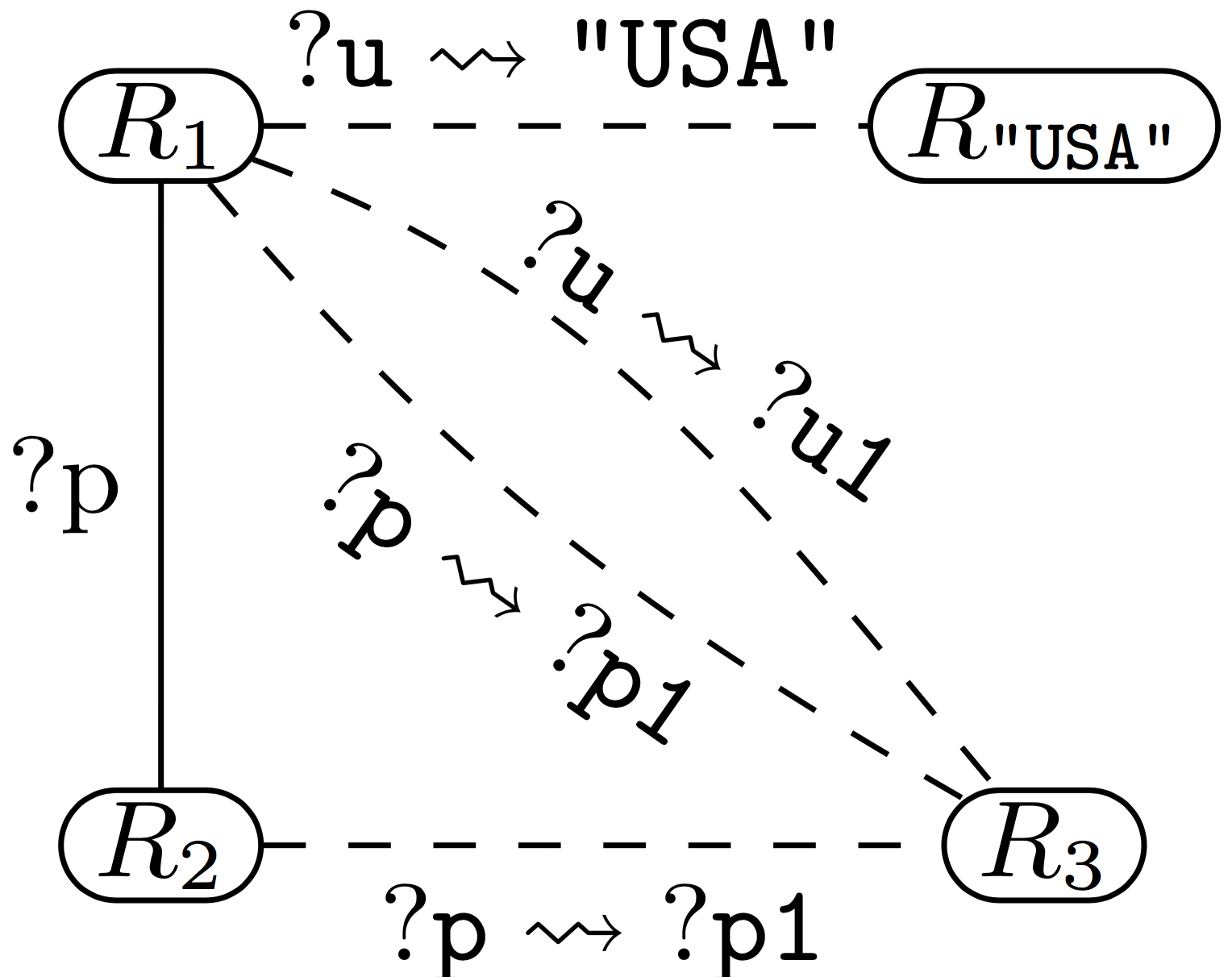}}
\vspace{-4mm}
\end{minipage}
\vspace{-2mm}
\caption{Example SwPP query and its graph representation}
\label{fig:query}
\vspace{-4mm}
\end{figure}

Figure~\ref{fig:query} shows an example SwPP query and its corresponding graph representation. Here, patterns $\mathtt{P_1}$, $\mathtt{P_2}$, $\mathtt{P_3}$ are basic triple patterns whose property each consists of a single URI. These are represented as vertices $R_1$, $R_2$, $R_3$, respectively, in the query graph. Since, the property path of $P_4$ points to only the constant {\tt "USA"}, a separate vertex $R_{\text{\texttt{"USA"}}}$ representing a relation that consists of just a singleton tuple is added to the query graph. The equi-join on the shared variable {\tt ?p} is represented by the continuous line between $R_1$ and $R_2$. A reachability edge, denoted by a dashed line for each property path, is added between the respective subjects' and objects' query vertices. This is the case between $R_1$ and $R_{\text{\texttt{"USA"}}}$ for the property path of $P_4$, between $R_1$ and $R_3$ for the property paths of $P_5$ and $P_6$, and between $R_2$ and $R_3$ for $P_5$.

\vspace{-4mm}
\subsection{Query Optimization}
\label{sec:optimization}
\vspace{-2mm}
Once the query is translated into its graph representation, classical join-order-enumeration techniques~\cite{triad,rdf3x10} can be employed to find a cost-efficient execution plan. We extend TriAD's optimizer to handle SwPP queries by adding a new operator---{\em Distributed Reachability Join} (DRJ)---and respective cost estimator for property paths. Next, we briefly discuss these operators, which is followed by a discussion of the cost estimation and join-order enumeration. 

\vspace{-4mm}
\subsubsection{Physical Query Operators.}
\label{sec:operators}
TriAD employs three physical operators---coined Distribu\-ted Index Scan (DIS), Distributed Merge Join (DMJ) and Distributed Hash Join (DHJ)---for processing index scans and equi-joins among triple patterns in SPARQL 1.0. Each of these operators works over the sharded partitions of the triple indexes described in Section~\ref{sec:tripleindexes} in parallel. In short, the DIS operators, which only occur at the leaves of the query plan, each build a relation by invoking a parallel scan over the respective SPO permutation index that was selected by the optimizer. The DMJ and DHJ operators each take two sharded relations plus the join keys (i.e., the shared variables) as input and perform a hash- or merge-join, respectively, to generate a new intermediate relation. 

\vspace{-4mm}
\subsubsection{Distributed Reachability Join (DRJ).}
\label{sec:drj} 
Analogously, we define a new DRJ operator to process triple patterns with property paths in TriAD*. This enhanced join operator takes two sharded relations $R_i$, $R_j$ as input and returns as output the subset of tuples in the cross-product $R_i \times R_j$, for which all of the attached {\em join conditions} $\mathcal{C}$ hold:
 \squishlist
 \item for each shared variable $\mathtt{?x}$ in $\mathcal{C}$, a pair of tuples in $R_i$ and $R_j$ must have equal values for $\mathtt{?x}$; and 
 \item for each reachability predicate $\mathtt{?x}\leadsto \mathtt{?y}$ in $\mathcal{C}$, a vertex $s$ that becomes bound to $\mathtt{?x}$ by a tuple in $R_i$ must be reachable to a vertex $t$ that becomes bound to $\mathtt{?y}$ by a tuple in $R_j$.
 \squishend
The evaluation of the DRJ operator is backed by the index structures for property paths described in Section~\ref{sec:swppindexes}.
That is, for sets of source and target vertices that become bound to the variables of a reachability predicate $\mathtt{?x}\leadsto\mathtt{?y}$, the DRJ operator performs a distributed set-reachability operation using the precomputed graph indexes over all partitions in parallel.

\vspace{-4mm}
 \subsubsection{Join-Order Optimization.}
 \label{sec:queryoptimization}
The DP table of the optimizer is initialized with the cost estimates for the DIS operations of each query vertex $R_i$. The scan costs for $R_i$ depend on whether the constants in the triple pattern match a respective SPO permutation index $\mathit{idx}$. For instance, if the subject and predicate are constants and the object is a variable, choosing an SPO or PSO permutation costs much less compared to any of the remaining permutations. 

In the query graph, we introduce two special kinds of query vertices, namely one for property paths with a constant subject or object, and one for property paths with (at least one) unbound variable. Scanning a singleton tuple as input has a unit cost of 1, while scanning a relation constructed for an unbound variable corresponds to the number of triples in the subgraph $G^p=(V^p, E^p)$ of $G_D$ that is induced by property $p$. In the latter case, we thus set the cardinality $\mathit{Card}(R_i)$ of a unary relation $R_i$ that is constructed from all vertices in $G^p$ to $|V^p|$. As an example, consider the property path {\tt ?x locIn* ?y}, and let variable $\mathtt{?y}$ be unbound (i.e., not occurring as a shared variable in any other triple pattern). Then the number of unique bindings for $\mathtt{?y}$ is the number of vertices in the induced subgraph consisting only of {\tt locIn} edges. Equation~\ref{eq:discost} summarizes the cost estimates we obtain for a DIS operator with respect to the precomputed cardinalities $\mathit{Card}(R_i)$ and available SPO permutations.
\begin{equation}
 \small
 \label{eq:discost}
 \mathit{Cost}(R_i) \propto
 \begin{cases}
	 1 & \text{if $R_i$ is a singleton tuple;} \\
	 \mathit{Card}(R^{\mathit{idx}}_i)/k & \text{if $R_i$ is sharded across $k$ slaves}\\
								& \text{and matches SPO index $\mathit{idx}$;}\\
	 \infty & \text{otherwise.}
 \end{cases}
\end{equation}

Once the DP table is initialized with the costs estimates for the DIS operators, we continue to build the join tree in a bottom-up manner. At each DP step, we merge two branches $Q^{left}$, $Q^{right}$ into a combined plan $Q$ by a join operator $\mathit{op}$ together with a set of join conditions $\mathcal{C}$. If there is at least one reachability edge between two relations $R_i$, $R_j$ that connect $Q^{left}$ and $Q^{right}$, a DRJ operator is employed. Note that each DRJ operator may invoke multiple distributed set-reachability queries depending on the number of edges that connect $R_i$ and $R_j$. That is, for the case $Q^{left} := \Join^{DMJ}(R_1, R_2)$ and $Q^{right} := R_3$ shown in Figure~\ref{fig:query}, we need the DRJ operator to consider two reachability predicates {\tt ?p workedWith* ?p1} and {\tt ?u sameState* ?u1}. Choosing the correct order for executing multiple reachability predicates thus is a sub-goal of overall optimization. 
Assuming independence among the join conditions $\mathcal{C}$, we plug in our precomputed index statistics as follows.
\begin{equation}
\small
\label{eq:joincost}
\mathit{Cost}(Q^{left}\Join^{op}_{\mathcal{C}}Q^{right}) \propto \sum_{C_i\in\mathcal{C}} \mathit{Card}(Q^{left}_{i}) \cdot \mathit{Card}(Q^{right}_{i}) \cdot \mathit{Sel}(C_i)
\end{equation}

While processing the conditions $C_i \in \mathcal{C}$, we also iteratively estimate the cardinality $Card(Q_i)$ of a subquery $Q_i$ of Equation~\ref{eq:joincost} as defined next. 
\begin{equation}
\small
Card(Q_i) := 
  \begin{cases}
    Card(Q_1) & \text{if $i=1$}\\
    \prod_{j=1}^{i-1}Card(Q_j)\cdot \mathit{Sel}(C_j) & \text{if $i > 1$}\\
  \end{cases}
\end{equation}

Thus, if $C_j$ is a graph-reachability predicate, $\mathit{Sel}(C_j)$ denotes the reachability selectivity $\mathit{Sel}(p)$ of the property $p$ that is associated with $C_j$. If $C_j$ refers to an equi-join, $\mathit{Sel}(C_i)$ denotes the precomputed join selectivity $\mathit{Sel}(p_i,p_j)$ for the pair of properties associated with the two triple patterns of the equi-join. 
The combined cost for a (sub-)query $Q$ then is defined recursively.
\begin{equation}
\small
\label{eq:combined}
  Cost(Q) = 
  \begin{cases}
    \max(Cost(Q^{left}), Cost(Q^{right}))\\
    +~Cost(Q^{left} \Join^{op}_{\mathcal{C}} Q^{right})\\
    +~Cost(Q^{left} \rightleftharpoons^{op} Q^{right})
  \end{cases}
\end{equation}

Here, $Cost(Q^{left} \Join^{op}_{\mathcal{C}} Q^{right})$ denotes the cost of processing the join operator $op \in \{\text{DMJ}, \text{DHJ}, \text{DRJ}\}$ with $Q^{left}$ and $Q^{right}$ as operands and join conditions $\mathcal{C}$ (Eq.~\ref{eq:joincost}). Likewise, $Cost$ $(Q^{left} \rightleftharpoons^{op} Q^{right})$ accounts for the shipping costs that incur when the resharding of intermediate relations is required. The shipping cost is proportional to the size and width of $Q^{left}$ and $Q^{right}$, respectively. Using $\max(\cdot,\cdot)$ as cost aggregation finally accounts for the parallel execution of the two branches~\cite{triad}. Figure~\ref{fig:qplan} shows an example query plan for the query of Figure~\ref{fig:query}.
\begin{figure}[t]
\centering
\vspace{-3mm}
\scalebox{0.24}{\includegraphics{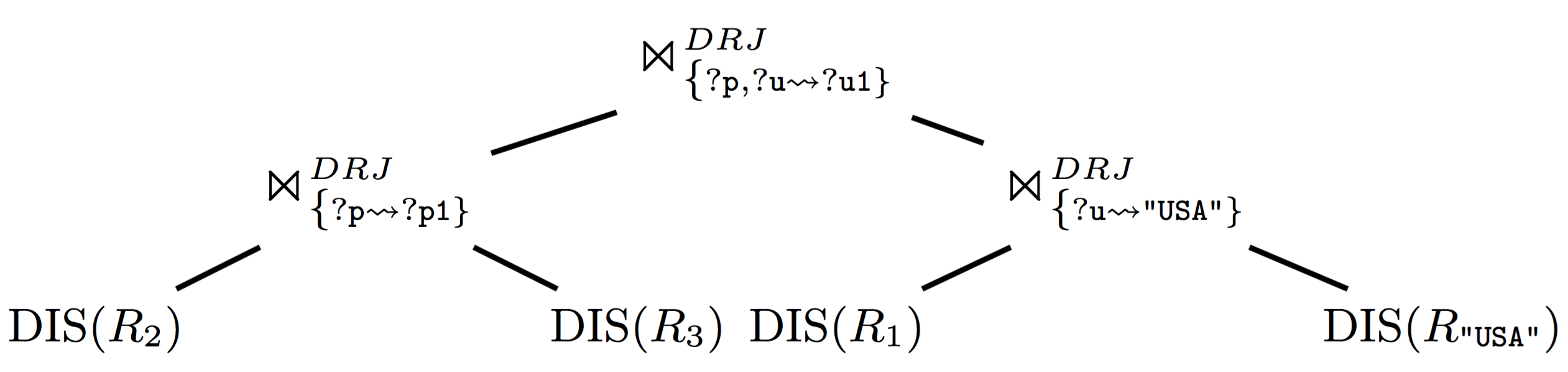}}
\caption{Example plan for the query of Figure~\ref{fig:query}}
\label{fig:qplan}
\vspace{-2mm}
\end{figure}

\vspace{-4mm}
\subsection{Distributed Query Processing}
\label{sec:execution}
\vspace{-2mm}
We embed the new DRJ operator into TriAD's multi-threaded and asynchronous processing framework to support the distributed execution SwPP queries. The principal processing flow and communication protocol~\cite{triad} remain unchanged and merely require an additional initialization of the source and target vertices for the distributed set-reachability queries, which are now triggered by the DRJ operators at their respective positions in the query plan.



\smallskip
{\bf 1. Scanning Base Relations.} The leaves of the operator tree always represent distributed index scans (DIS). Each slave scans its local SPO permutation index and selects tuples according to the constants associated with the DIS operator. Due to the layout of our SPO indexes, this merely requires initializing an iterator at the first tuple in a permutation list that matches the constants.
For a DRJ operator with a reachability predicate, whose source or target is a single constant, a singleton relation is created directly from that constant. If the DRJ operator has a reachability predicate with an unbound variable via a property $p$, a (sharded) unary relation with the local vertices of $V^p$ is created. 

\smallskip
{\bf 2. Query-Time Sharding.} During the execution of the query plan, resharding of intermediate relations may be required to ensure the proper execution of joins (DMJ, DHJ) and set-reachability (DRJ) operations. With six SPO permutations, each DMJ operator requires sharding of at most one of its base relations at query time, while the DHJ operator requires sharding of at least one of its intermediate relations, depending on the locality of the tuples with respect to the join key. Sharding for the DRJ operator depends on the locality of the join keys based on shared variables (if present) and the locality of the vertices that become bound to the source and target variables of the reachability predicates. Thus, resharding may be required for both input relations of a DRJ operator.

\smallskip 
{\bf 3. Parallel Execution of Operators.} In addition to the concurrent execution of the operators across the slaves, each slave also locally pursues the execution of the query plan in a multi-threaded fashion. Starting from the leaves of the query plan, all operators are locally executed in one separate thread for each {\em execution path} (EP) (i.e., for each distinct leaf-to-root path) in the query plan. Since slaves may take different amounts of time to execute an operator over their local partition of the index, an {\em asynchronous} exchange of messages for resharding the partial relations at query time makes this step more efficient than a synchronous protocol. As soon as all the shards for the two input relations of a join operator are in place, the threads of the two EPs at each slave are merged into one, and the next join operations can be invoked locally. 
For a DRJ operator with a graph-reachability predicate, whose source or target variables become bound to constants due to a shared variable, the respective source and target sets for the distributed set-reachability query are initialized from those constants. These are then resharded to the slaves that hold the graph partitions containing the source and target vertices.

\vspace{-2mm}

\vspace{-2mm}
\section{Evaluation}
\label{sec:experiments}
\vspace{-2mm}
TriAD* is entirely implemented in C++. We used GCC-4.7.3 with {\tt -O3} optimization and MPICH2-1.4.1 and Boost-1.55 as external libraries. We ran all of the following experiments on a compute cluster with up to 11 nodes, out of which 1 was dedicated as the master node. Each node runs Debian 7.5, has 48GB of RAM and an Intel E5530@2.40GHz quad core CPU with HT enabled.

\smallskip
\noindent {\bf Datasets.} We used three large-scale, both real-world and synthetic, RDF datasets for our evaluation: (i) LUBM-500M\footnote{\url{http://swat.cse.lehigh.edu/projects/lubm/}}  (scaled to 500 million triples) is generated using UBA 1.7 in N3 format, (ii) Freebase-500M (with 500 million triples) refers to a subset of a recent Freebase snapshot\footnote{\url{https://developers.google.com/freebase/data}} and (iii) a recent snapshot of DBpedia\footnote{\url{http://downloads.dbpedia.org/2015-04/core/}} (with 417,445,957 triples).

\smallskip
\noindent {\bf Queries.} 
We manually designed three queries for each dataset (L1--L3 for LUBM, F1--F3 for Freebase, D1--D3 for DBpedia) to capture a mixture of reachability queries and relational joins. All queries are listed in our Appendix.



\vspace{-4mm}
\subsection{Distributed SwPP Queries}
\label{sec:swpp-exp}
\vspace{-2mm}
We first discuss the distributed processing of SwPP queries for the fixed snapshot of the three datasets described above. 
For TriAD*, we used 5 slaves for this setting (plus 1 master node). 
As competitor, we used the Virtuoso 7.1.0 native RDF store, which is the only available RDF store we are aware of that supports full property-path processing. We remark that the open-source edition of Virtuoso 7.1.0 does not support distribution. We thus compare against a centralized installation of Virtuoso on one of our compute nodes.
\begin{table}[htb]
\vspace{-2mm}
  \fontsize{7.2}{7.5}\selectfont
  \centering
  \tabcolsep=0.23cm
  \begin{tabular}{|lrrrrr|}
    \hline
    \multicolumn{6}{|c|}{(a) LUBM-500 \it (query times in seconds)}\\
    &{\bf \#Slaves}&{\bf L1}&{\bf L2}&{\bf L3}&{\bf Geo.-Mean}\\
    \hline
    TriAD*&1&6.437&0.331&42.681&4.497\\
    TriAD*&5&{\bf 1.250}&{\bf 0.162}&{\bf 8.516}&{\bf 1.199}\\
    Virtuoso (cold)&1&10.050&12.624&57.776&19.425\\
    Virtuoso (warm)&1&4.963&5.452&56.603&11.527\\
    \hline
    \hline
    \multicolumn{6}{|c|}{(b) Freebase-500 \it (query times in seconds)}\\
    &{\bf \#Slaves}&{\bf F1}&{\bf F2}&{\bf F3}&{\bf Geo.-Mean}\\
    \hline
    TriAD*&1&1.084&1.568&0.677&1.048\\
    TriAD*&5&{\bf 0.356}&0.642&{\bf 0.423}&0.459\\
    Virtuoso (cold)&1&6.590&4.112&13.809&7.206\\
    Virtuoso (warm)&1&1.196&{\bf 0.002}&5.601&{\bf 0.238}\\
    \hline
    \hline
    \multicolumn{6}{|c|}{(c) DBpedia \it (query times in seconds)}\\
    &{\bf \#Slaves}&{\bf D1}&{\bf D2}&{\bf D3}&{\bf Geo.-Mean}\\
    \hline
    TriAD*&1&24.822&0.713&29.407&8.044\\
    TriAD*&5&{\bf 7.973}&{\bf 0.412}&{\bf 11.223}&{\bf 3.328}\\
    Virtuoso (cold)&1&46.185&19.352&317.899&65.741\\
    Virtuoso (warm)&1&27.820&2.395&302.753&{27.222}\\
    \hline
  \end{tabular}
  \vspace{2mm}
 \caption{Performance evaluation of SwPP queries}
  \label{tab:swppeval}
 \vspace{-8mm}  
\end{table}

\smallskip
{\bf A. LUBM-500M.} The results for processing SwPP queries (L1, L2, L3) are shown in Table~\ref{tab:swppeval}(a).  L1 resembles a single, non-selective reachability join. Processing L1 thus involves an index scan for two input relations and a respective evaluation of the reachability join. We can observe that the centralized version of TriAD* performs better than Virtouso in a {\em cold} cache and comparable to Virtuoso in a {\em warm} cache setting. We however achieve a significant scale-out for L1 when we evaluate the query on a cluster of 5 slaves. Next, L2 is a selective query with two regular joins and a single reachability join. For this query, we can observe that TriAD* achieves a better performance compared to Virtuoso in both the cold and warm cache settings. The non-selective query L3 contains two reachability joins in conjunction with two regular joins. Also here, TriAD* continues to perform better than Virtuoso under both a cold and warm cache and further scales out very well in a distributed setting.

\smallskip
{\bf B. Freebase-500M.} For Freebase, we considered three queries (F1, F2, F3) which we designed along the lines of the L1, L2, L3 LUBM queries. The performance of TriAD* for Freebase-500M shows a similar behavior as the one we observed for LUBM-500M. The results are shown in Table~\ref{tab:swppeval}(b). F1 again consists of a single, non-selective reachability join. For this query, TriAD* performs better than Virtuoso under a {\em cold} cache and has comparable performance to Virtuoso in a {\em warm} cache setting. For the selective query F2, which comprises of regular joins and a single reachability join, TriAD* performs better than Virtuoso in the cold-cache setting, but Virtuoso with a warm cache outperforms TriAD* in the both centralized and distributed settings. For the non-selective query F3, which comprises of multiple reachability joins along with regular joins, TriAD* performs significantly better than Virtuoso under both a cold and warm cache. 
We remark that, for F3, Virtuoso tends to report different results over repeated runs of the query, which indicates problems with their current support for property paths.

\smallskip
{\bf C. DBpedia.} We once more considered three queries (D1, D2, D3) consisting of a mixture of relational joins and graph-reachability predicates for DBpedia. The runtime performance of TriAD* in comparison with Virtuoso is shown in Table~\ref{tab:swppeval}(c). Also here, TriAD* continues to perform very well compared to Virtuoso under both cold and warm cache settings.  

\vspace{-4mm}
\subsection{Scalability Tests}
\label{sec:scal-exp}
\vspace{-2mm}
We finally evaluated the scalability of TriAD* for SwPP queries by varying the number of slaves from 1 to 10. For this evaluation, we again considered LUBM-500M, Freebase-500M and DBpedia.  The results under {\em strong scaling} are shown in Figure~\ref{fig:rscale}(a) for LUBM-500M, in Figure~\ref{fig:rscale}(b) for Freebase-500M, and in Figure~\ref{fig:rscale}(c) for DBpedia, respectively. 
As our last series of runs, we also evaluated the performance of TriAD* under {\em weak scaling}, by increasing the size (from 20\%--100\%) of the collections as well as the number of slaves (from 2--10) in equal proportions. The results are shown in Figure~\ref{fig:rscale}(d)--(f).
 \begin{figure*}[htb]
  \centering
   \vspace{-2mm}
   \scalebox{.31}{\includegraphics{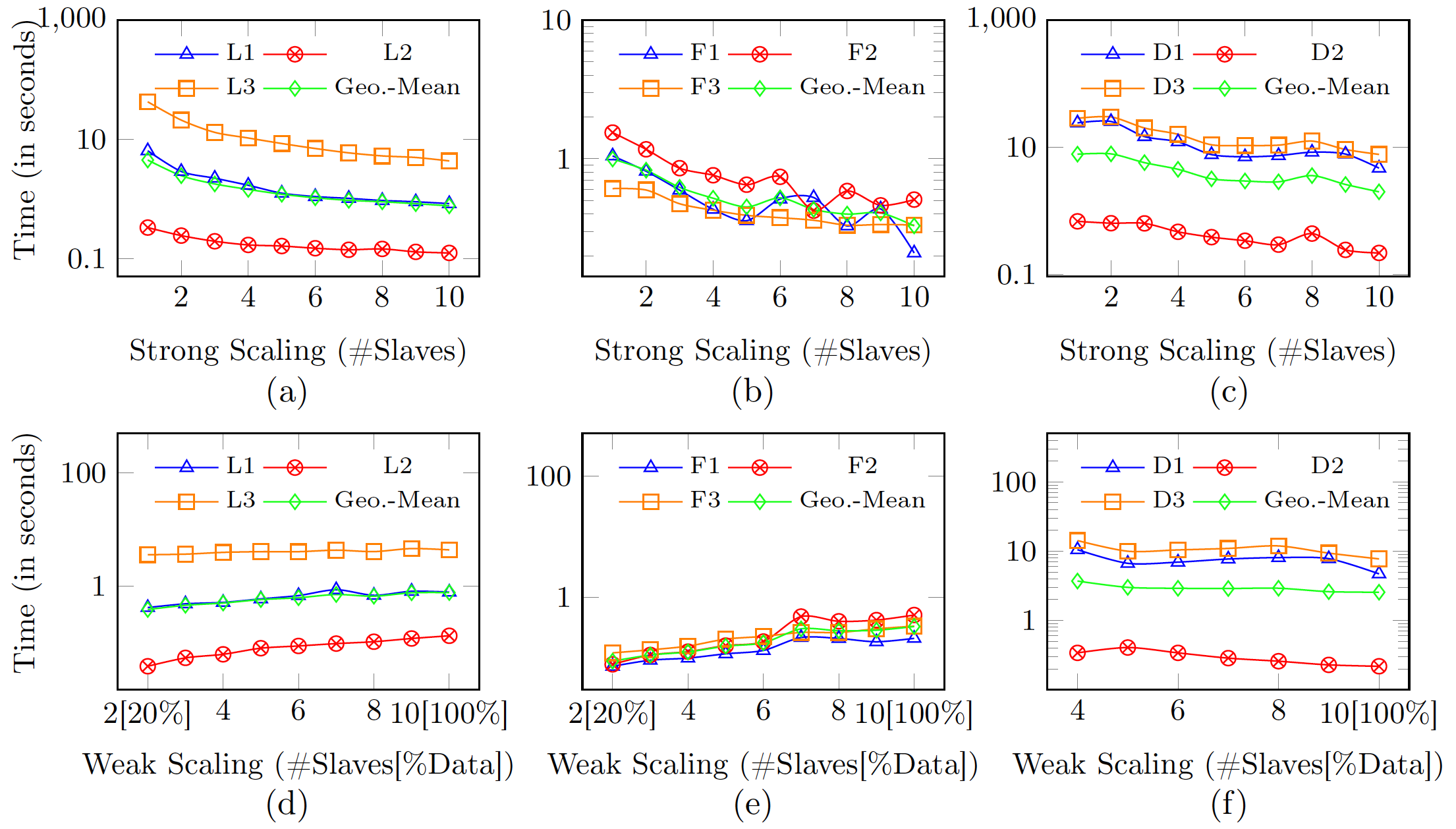}}
   \vspace{-2mm}
   \caption{Strong (a)--(c) and weak (d)--(f) scalability of SwPP queries}
   \label{fig:rscale}
   \vspace{-4mm}
\end{figure*}


\section{Conclusions}
\label{sec:conclusions}
\vspace{-4mm}
We presented the architecture of TriAD*, which to our knowledge is the currently fastest, distributed RDF engine that explicitly tackles the processing of property paths according the recently updated SPARQL 1.1 specification. Building on top of our TriAD engine, we leverage its multi-threaded and asynchronous query processing framework to implement a new relational query operator to tackle the kind of generalized graph-pattern queries that arise in SPARQL 1.1.
Our evaluation over both real-world and synthetic RDF collections confirm that TriAD* achieves very significant gains compared to the only currently available, native RDF store that supports SPARQL 1.1 with property paths. As for future work, we intend to extend TriAD* to support also more complex property-path variants, including paths with length restrictions, shortest paths, as well as paths with more general regular expressions. 
\vspace{-2mm}



{ \small
\bibliographystyle{abbrv}
\bibliography{main}  
}

\newpage
\appendix

\section{SwPP Queries}
\label{app:sparqlqueries}
\vspace{-2mm}
\subsection{LUBM Queries}
\vspace{-2mm}
{\scriptsize 
\noindent 
{\tt @prefix rdf: <http://www.w3.org/1999/02/22-rdf-syntax-ns\#>}\\
{\tt @prefix ub: <http://www.lehigh.edu/$\sim$zhp2/2004/0401/univ-bench.owl\#>}

\smallskip
\noindent{\tt \bf L1:} {\tt SELECT * WHERE \{ ?x rdf:type  ub:ResearchGroup .} {\tt ?x ub:subOrganizationOf* ?y.}\\$~~~~~~${\tt ?y rdf:type ub: University. \}}

\smallskip
\noindent{\tt \bf L2:} {\tt SELECT * WHERE \{ ?x rdf:type  ub:FullProfessor. ?x ub:headOf ?d.}\\$~~~~~~${\tt ?d ub:subOrganizationOf* ?y. ?y rdf:type ub:University.\}}

\smallskip
\noindent{\tt \bf L3:} {\tt SELECT * WHERE \{ ?r1 rdf:type  ub:ResearchGroup .} {\tt ?r1 ub:subOrganizationOf* ?y.}\\ $~~~~~~${\tt ?y rdf:type ub:University . ?r2 rdf:type  ub:ResearchGroup. ?r2 ub:subOrganizationOf* ?y.\}}
}

\subsection{Freebase Queries}
\vspace{-2mm}
{\scriptsize
\noindent 
{\tt @prefix rdf: <http://www.w3.org/1999/02/22-rdf-syntax-ns\#>}\\
{\tt @prefix fb: <http://rdf.freebase.com/ns>}

\smallskip
\noindent{\tt \bf F1:} {\tt SELECT * WHERE \{ ?p fb:people.person.place\_of\_birth ?city . \\$~~~~~$?city fb:location.location.con\-tainedby* ?state. \\ $~~~~~$?country fb:location.location.contains ?state. \}}

\smallskip
\noindent{\bf F2:} {\tt SELECT * WHERE \{ ?p fb:people.person.place\_of\_birth ?city .  \\ $~~~~~$?city  fb:location.location.con\-tainedby* ?state.  ?country fb:location.location.contains\\ $~~~~~$?state. ?p fb:award.award \_winner.awa\-rds\_won ?prize. \\ $~~~~~$?p rdf:type  fb:government.us\_president. \}}

\smallskip
\noindent{\bf F3:} {\tt SELECT * WHERE \{?p fb:award.award\_winner.awards\_won ?prize. ?prize rdf:type* ?z . \\ $~~~~~$?z fb:aw\-ard.award\_honor.ceremony> ?c.?p fb:people.person.sibling\_s* ?p1. \\ $~~~~~$?p1 fb:award.award\_winner.awa\-rds\_won ?prize. \}}
}

\subsection{DBpedia Queries}
\vspace{-2mm}
{\scriptsize
Namespace prefixes available at: \url{http://de.dbpedia.org/sparql?nsdecl}

\smallskip
\noindent {\bf D1:} {\tt SELECT * WHERE \{ ?s1 rdf:type ?s. ?s rdfs:subClassOf* ?o. \\ $~~~~~$?o owl:equivalentClass yago-res:word\-net\_medium\_106254669 \}}

\smallskip
\noindent {\bf D2:} {\tt SELECT * WHERE \{ ?s foaf:isPrimaryTopicOf wiki:North\_Auburn,\_California . \\ $~~~~~$?s dbpedia-owl:isPartOf* ?c. ?x dbpedia-owl:hometown ?c. ?x foaf:isPrimaryTopicOf ?r. \}}

\smallskip
\noindent {\bf D3:} {\tt SELECT * WHERE \{ ?s dbpprop:leaderTitle ?title. ?title rdf:type ?class. \\ $~~~~~$?class rdfs:subClassOf* ?class2. \\ $~~~~~$?class2 owl:equivalentClass yago-res:wordnet\_abstraction\_100002137 . \\ $~~~~~$?s foaf:isPrimaryTopicOf wiki:North\_Auburn,\_California . ?s dbpedia-owl:isPartOf* ?c.\\ $~~~~~$?x dbpedia-owl:hometown ?c. ?x foaf:isPrimaryTopicOf ?r. \} }

}


\end{document}